\documentclass[11pt]{article}
\usepackage{times}
\usepackage{graphicx}
\usepackage{color}
\usepackage{amssymb}
\usepackage{amsmath}
\usepackage{ifthen}
\usepackage{calc}
\usepackage{multirow}

\newtheorem{theorem}{Theorem}
\newtheorem{corollary}[theorem]{Corollary}

\newcommand{\nlog}[2]{\ensuremath{\Theta\left(n\ifthenelse{\equal{#1}{1}}{}{^{#1}}\ifthenelse{\equal{#2}{0}}{}{\log\ifthenelse{\equal{#2}{1}}{}{^{#2}} n}\right)}}

\newlength{\Ainlength}
\newlength{\Aintemp}

\begin{document}

\begin{center}
{\Large \textbf{Fine-Grained Computation in 3-Space:\\ Matrix Multiplication and Graph Problems}}
\bigskip \bigskip

{\large Quentin F. Stout}\\[\medskipamount]
Computer Science and Engineering, University of Michigan
\end{center}
\bigskip

\begin{abstract}
Obeying constraints imposed by classical physics, we give optimal fine-grained algorithms for matrix multiplication and problems involving graphs and mazes, where all calculations are done in 3-dimensional space.
We assume that whatever the technology is, a bit requires a minimum volume and communication travels at a bounded speed. 
These imply that multiplying $n \times n$ matrices takes $\Omega(n^{2/3})$ time, and we show that this can be achieved by a fine-grained 3-d mesh of $n^2$ processors.
While the constants are impractically large, this is asymptotically faster than parallel implementations of Strassen's algorithm, while the lower bound shows that some claims about parallelizing faster serial algorithms are impossible in 3-space.
If the matrices are not over a ring then multiplication can be done in $\Theta(n^{3/4})$ time by expanding to a mesh larger than the input.
In 2-d (such as the surface of a chip) this approach is useless and $\Theta(n)$ systolic algorithms are optimal even when the matrices are over a ring.
Similarly, for path and maze problems there are approaches useful in 3-d but not 2-d.

\end{abstract}
\medskip


\section{Introduction}

We are interested in understanding the algorithmic differences in fine-grained parallel computing for computation in 3-dimensional space vs.\  the extensively studied 2-dimensional space used in VLSI, and for a few of the differences between medium-grained systems vs.\ fine-grained ones.
Whatever the technology being used, we assume that a word of data takes a unit volume or area, respectively, and, for fixed energy takes unit time to move a unit distance.
These limits are imposed by classical (non-quantum) physics, and imply that the basic mesh-connected computer is the densest packing of computation.
This is in contrast to PRAM models that ignore the physical implications of increasing the memory and treat the communication time as being constant no matter what the distance.
It is also in contrast to serial computer models where it is assumed that the time to access RAM is independent of problem size, and in contrast to medium or large-grained distributed memory parallelism models which make the RAM assumption on a board and typically make a PRAM-like assumption of constant time communication time between boards.

Throughout, our computer is a 3-dimensional mesh (2-d meshes will also be discussed), with each processor connected to the 6 adjacent processors.
The processing will be done in non-overlapping subcubes.
Each time step is actually split into two smaller time steps, one for every processor to pass information to its neighbors, if it has any communication to do, and the other to do its processing as a member of a subcube (if it is in an active subcube).
The time to move a word of information from one processor to another is proportional to the $L_1$ distance between them if a shortest path is used, and more generally is linear in the length of the path.

One aspect of the mesh model is that while communication is relatively slow, it eliminates the volume or area of wires to increase the bandwidth from one subcube to another.
An algorithm which increases the number of long wires as $n$ increases will increase the total volume.
This was recognized as a serious issue in VLSI layout where typically the space increase was considered, making many of the wires longer, while the increased time to travel on a wire was occasionally, though not always, ignored.
The space increase alone made some desirable configurations vastly inferior than if wires had zero area~\cite{Ullman} since it required chips that had more area than the area needed to do the operations between words.
An important example is the binary hypercube~\cite{LeisersonHypercube}:
\begin{quote}
A hypercube, which is a popular
interconnection network for parallel computers, requires considerably more
area --- $\Theta(n^2)$.
What causes a hypercube to occupy so much area? Although the size of a vertex grows slowly with the number of vertices in a hypercube, most of the area of a hypercube layout is devoted to wires.
\end{quote}
Here $n$ is the number vertices, not the number of rows or columns in the matrices considered in this paper, hence for our purposes the area is $\Theta(n^4)$

Largely motivated by VLSI, there is extensive literature on algorithms for the fine-grained 2-dimensional mesh (see~\cite{Lei92,MIT,Ullman}).
A few fine-grained algorithms, and questions, for the 3-dimensional mesh had appeared, such as for sorting~\cite{ThKuSort} (including finding the optimal high-order coefficient of time~\cite{KundeSort,Suel}),
oblivious routing~\cite{IwamaOblivious3d}, connected component labeling of graphs~\cite{QFSMST84}, and reachability in mazes~\cite{Beyer,perceptrons,QFSmaze}, but overall there are far fewer algorithms than for 2-d.

However, interest in 3-dimensional computing is mounting due to its potential for faster and more energy efficient data movement~\cite{Pavidis3D,Topol3D}.
It lowers the communication diameter and increases the bisection bandwidth, fundamental lower bounds for many problems.
It was an important step towards reaching exascale~\cite{DARPA3D,DOE:SciGrandChal,DARPAExascale},
and now must be used to reach zetascale.
One saying is that ``flops are free'', meaning that the time and energy required to move data on a chip from L2 cache to the floating point units are greater than doing a few calculations.
Further, moving from RAM onto the chip is far more costly in time and energy.
3-d systems can simultaneously reduce time and energy.

In the `80s there was theoretical work on embedding 2-d circuits into 3-d layouts, using the 3rd dimension to make denser devices on the base layer, with wiring in the higher layers.
In some cases devices were allowed to be on higher layers as well.
When the 3rd dimension could have the same extent as the others, with results such as Leighton and Rosenberg's theorem that the smallest volume of a 3-dimensional layout of an N-device circuit is no more than $\sqrt{NA}$, where A is the smallest area of a 2-d layout~\cite{LeRo3dcircuit}.
Another result is that every small-degree $N$ vertex graph can be laid out in a 3-dimensional grid with volume $\Theta(N^{3/2})$ and wire-length $\Theta(N^{1/2})$~\cite{Rosenberg3DIC}, in contrast to the 2-d bounds of area $\Theta(N^2)$ and wire length $\Theta(N)$~\cite{Leightonlayout82}.
These results depend on having the third dimension grow as a power of $N$.
All of these 3-d embeddings can be used in our model, but we are more interested in developing algorithmic techniques.

This raises questions as to what algorithmic opportunities appear when moving from 2 to 3 dimensions.
Some are trivial extensions and quite straightforward.
For example, algorithms for sorting~\cite{ThKuSort} and component labeling of a graph given its edges~\cite{QFSMST84} easily extend to all dimensions, with the time going from $\Theta(\sqrt{n})$ on 2-d meshes to $\Theta(\sqrt[d]{n})$ in d-dimensional meshes of $n$ processors with one data item per processor and equal extents in all dimensions
(the implied constants in the O-notation depend on $d$\,).
Often sorting (or permuting data) is the time constraint on divide-and-conquer mesh algorithms, thus many easily extend to higher dimensions with the same times as sorting (e.g., see~\cite{MIT}).
For some problems there are subtle changes, e.g., oblivious  and standard routing take the same time in 2-d,
but are $\Theta(\sqrt{n})$ vs.\ $\Theta(\sqrt[3]{n})$ in 3-d~\cite{IwamaOblivious3d,OsterlohOblivious}.

Here we consider problems where significant changes occur: matrix multiplication and corollaries such as determinant~\cite{Strassen}; all-pairs shortest paths; and finding shortest paths in a maze.
Their 2-d versions been studied for decades, but 3-d have not.
Our emphasis is on showing that moving from 2-d to 3-d can exploit new approaches, such as 
embedding the problem into a larger mesh.
See Table~\ref{tab:Improvements}.

\section{Mesh Model} \label{sec:basic}

Our computational model is a mesh-connected computer where each processor is connected to its 6 (in 3-d) or 4 (in 2-d) adjacent neighbors.
Computations and exchanging a word of information with a neighbor take unit time.
Meshes with $p$ processors have a fixed number of words of $\Omega(\log p)$ bits, enough to allow processors to store their coordinates.
We slightly abuse notation and count words as unit volume and operations on words take unit time, i.e., we're ignoring a log factor in both to better match normal analyses.
In contrast, cellular automata have only $\Theta(1)$ bits of memory, hence typically analyzing their volume and time does not hide such a factor but it greatly complicates the algorithms.
This is further discussed in Section~\ref{sec:final}.

Our 2-d meshes are always square, and 3-d always cubical.
Thus for both the communication diameter is linear in the edgelength.
In meshes larger than the input size the input always starts in a corner, and the time of movement through the larger mesh can become a dominant factor.
The \textit{size} of a mesh is the total number of processors, and for a matrix is the total number of entries.

\subsection{Some Relevant Previous Results}

\noindent
\textbf{Sorting, Divide and Conquer:~~} For a d-dimensional mesh of size $N$, with an input of size $N$, Thompson and Kung~\cite{ThKuSort} showed that sorting can be done in $\Theta(N^{1/d})$ time.
Sorting can be used to support a variety of data rearrangement tasks, and in particular is useful for divide and conquer.
Using divide and conquer, finding a minimal spanning tree problem and component labeling can also be solved in $\Theta(N^{1/d})$ time when the input is an unordered list of edges~\cite{QFSMST84}.
Many other divide and conquer algorithms have similar time behavior.
In all of these algorithms the implied constants are a function of $d$.
\bigskip

\noindent
\textbf{3-d Simulation of 2-d:~~}
Even though square 2-d grid graphs cannot be embedded into cubical 3-d grids with constant dilation,
any algorithm on a 2-d mesh of size $N$ can be stepwise simulated on a 3-d mesh of size $N$ at constant amortized time per step.
To do so, divide the 2-d mesh into $s=N^{1/3}$ subsquares of edgelength $s$ and stack them into a cube (ordering is irrelevant).
The simulation has major time steps of length $s$.
If a layer of the 3-d mesh has subsquare $S$, in each major time step it simulates $S$ and its 8 neighboring subsquares.
It continually shrinks the region simulated, so that after $s$ time steps it is only simulating $S$. Then it gets the updated values being stored in $S$'s neighbors and starts the next major time step
(i.e., it is using ghost cells).
The entries marked ``sim'' in Table \ref{tab:Improvements} are based on this approach.

\section{Matrix Operations} \label{sec:mult}

We consider the matrix product $C = A \times B$ of $n \times n$ matrices $A$ and $B$, stored no more than one entry of each per processor, initially without any entry duplicated.
By \textit{general matrix multiplication} we mean that the terms are of the form
\small{$$C(i,j) = \bigoplus_{k=0}^{n-1} A(i,k) \otimes B(k,j) \mathrm{~~for~} \leq i, j \leq n-1$$}
\noindent
where $\oplus$ is an associative operation.
For timing analyses we assume the operations can be computed in unit time.
An algorithm for general matrix multiplication must be correct no matter what $\oplus$ and $\otimes$ are other than the associativity assumption (some authors require that $\oplus$ has an identity but this can be eliminated by just adding an artificial identity).
While faster algorithms are known for certain combinations of $\oplus$ and $\otimes$ (such as regular multiplication of matrices with real entries), we assume there are no algorithms faster than the straightforward one for arbitrary $\oplus$ and $\otimes$ combinations, and hence the time of any serial algorithm is $\Omega(n^3)$.
The standard serial implementation, taking $\Theta(n^3)$ time, is 

\begin{quote}
\texttt{\small
\hspace*{0.1in}initialize all entries of C to 0\\
\hspace*{0.1in}forall i,\! j $\in$ \{0,\!n-\!1\}\\
\hspace*{0.22in} for k=0 to n-1\\
\hspace*{0.45in}C(i,j) = C(i,j)\,$\oplus$\,(A(i,k)\,$\otimes$\,B(k,j))\\
} 
\end{quote}

By \textit{matrix multiplication over a ring} we mean that the $\oplus$ operation is over a ring that has an inverse (subtraction in the standard numerical case). 
While this is an instance of general matrix multiplication, the $\oplus$ inverse allows
approaches such as Strassen's matrix multiplication to be applied and hence $o(n^3)$ can be achieved.
We use $(\oplus,\otimes)$ to denote matrix multiplication, and always specify if it is general or over a ring.

Another problem which often stated in a form that requires computing the entries of a matrix is transitive closure, where $C(i,j)$ is a boolean variable which is true iff $i$ is a predecessor of $j$. 
The serial Floyd-Warshall algorithm, taking $\Theta(n^3)$ time, is
\begin{quote}
\noindent  \ \\
\texttt{\small
\hspace*{0.1in}initialize C = C$_0$\\
\hspace*{0.1in}for k= 0 to n-1\\
\hspace*{0.25in}forall i,\! j $\in$ \{0,\!n-\!1\}\\
\hspace*{0.4in}C(i,j) = C(i,j)\,$\oplus$\,(C(i,k)\,$\otimes$\,C(k,j))\\
} 
\end{quote}
\noindent
where $\oplus$ is $\vee$ and $\otimes$ is $\wedge$.
We denote this as a $(\oplus,\otimes)$ \textit{F-W matrix operation}.

\begin{table*}
\begin{center}

\begin{tabular}{|c|c|c|c|} \hline
Mesh     & Sort, MST & Matrix mult & Matrix mult \\ 
dim      & $n^2$ items     & $n \times n$     & $n \times n$   \\ 
         &    & general & over ring  \\ \hline
2-d     & $\Theta(n)$ \cite{QFSMST84,ThKuSort} & $\Theta(n)$ 
& $\Theta(n)$ \\ \hline
\multirow{2}{*}{3-d~}
prev. & $\Theta(n^{2/3})$ \cite{QFSMST84,ThKuSort} & $\Theta(n)$~ sim &  $\Theta(n)$~ sim \\ 
~~~~~~here & same & $\Theta(n^{3/4})$ & $\Theta(n^{2/3})$  \\ \hline
\end{tabular}

\medskip
sim: block-wise simulation of 2-d mesh\\
\medskip

\caption{Meshes have $\Theta(n^2)$ processors, except for}
\centerline{general matrix mult.\ on 3-d mesh ``here'', where there are $\Theta(n^{9/4})$} \label{tab:Improvements}

\end{center}

\vspace*{-0.1in}
\hrulefill
\end{table*}

\subsection{Matrix Operations in 2-Space} \label{sec:2dmult}

Well-known simple systolic algorithms solve general matrix multiplication on a 2-d torus in $\Theta(n)$ time. 
To compute $C = A \times B$, $C(i,j)$ is computed at processor $P(i,j)$.
This processor acts as an accumulator.
$A(i,k)$ and $B(k,j)$ arrive at $P(i,j)$ at time $(i+j+k) \mod n$ and are multiplied and added to the accumulator.
A torus can be flattened onto a 2-d mesh with dilation 2, so the mesh can simulate this in $\Theta(n)$ time.

Gentleman~\cite{Gentleman} made a simple observation: this algorithm is optimal up to multiplicative constants.
For any $A(i,j)$ let $B(k,\ell)$ be an element of $B$ of maximal distance from $A(i,j)$.
Since we have a bound on the density of words the distance between $A(i,j)$  and $B(k,\ell)$ is $\Omega(n)$ no matter how the entries are arranged.
The value of $C(i,\ell)$ depends on both of these values, and hence cannot be determined in time less than its maximum distance from either, which is $\Omega(n)$.
Thus in 2-d nothing clever is useful: not Strassen's multiplication, not rearranging entries, not adding more processors, etc.
There are a great many other problems which have a similar lower bound in 2-d.

For the $(\vee,\wedge)$ F-W matrix operation, van Scoy~\cite{VanScoy} gave a simple algorithm to finding the transitive closure in $\Theta(n)$ time on an $n \times n$ 2-d cellular automaton.
Her algorithm extends to general F-W operations, and by using 2-d meshes instead of cellular automata, 
in $\Theta(n)$ time a $(\min, +)$  F-W matrix operation implements Floyd's all pairs minimum distance algorithm and, by keeping track of the minimizing $k$ for each $C(i,j)$, determines all pairs shortest path (APSP) in the same time.
Here too the problem requires $\Omega(n)$ time.

\subsection{Matrix Operations in 3-Space} \label{sec:3dmult}

In 3-space a lower bound for operations on $n \times n$ matrices is $\Omega(n^{2/3})$ since that is the communication diameter of the smallest mesh able to hold the matrix, and Gentleman's argument applies again for matrix multiplication and many other problems.
For general matrix multiplication this is less than the lower bound obtained from linear speedup ($\Theta(n)$), i.e., it is impossible to achieve.
If instead one starts with the matrix values in an $n \times n$ corner of a larger mesh, where the initial values at other processors are irrelevant, then there are more processors and the linear speedup bound can be improved, though the communication diameter increases.
Once it is large enough that these two bounds are the same the optimum is achieved.

\begin{theorem} \label{thm:3dgenmult}
For any $\alpha$, $2 \leq \alpha \leq 9/4$, given two $n \times n$ matrices in a 3-dimensional mesh-connected computer of size $n^\alpha$, Algorithm A computes their general matrix product in $\Theta(n^{3-\alpha})$ time, where the implied constants depend on $\alpha$.
When $\alpha=9/4$ the time is $\Theta(n^{3/4})$, which is optimal for general matrix multiplication on a 3-dimensional mesh of any size.
\end{theorem}
\noindent \textit{Proof:}\,
The mesh has edgelength $n^{\alpha/3}$, and thus in step 1 a submatrix travels distance $O(n^{\alpha/3})$.
The recursive steps keep reducing this by a factor of 2, and hence the worst case total distance, and time to move that far, is $\Theta(n^{\alpha/3})$.
Step 4 takes a similar amount of time since combining the results is merely moving two submatrices together and doing termwise $\oplus$.
As noted in Section~\ref{sec:2dmult}, step 3 takes $\Theta(n^{3-c})$ time.

Optimality follows from the observations that linear speedup implies time is $\Omega(n^{3-\alpha})$, and having data reach $\Theta(n^\alpha)$ processors takes $\Omega(n^{\alpha/3})$ time.
Choosing $\alpha = 9/4$ minimizes the larger of these two.
$\Box$

\begin{figure}
\noindent General matrix multiplication of $n \times n$ matrices $C = A \times B$\\
in a 3-d mesh of size $n^\alpha$, $2 \leq \alpha \leq 9/4$.
\medskip

\begin{enumerate}
\item Let $A, B$ have quadrants $A_{i,j}$ and $B_{i,j}$, respectively, where $i,j \in \{0,1\}$,
and let the mesh have octants $O_{a,b,c}$ where $a,b,c \in \{0,1\}$.
Partition the problem into 8 subproblems of multiplying submatrices of size $\frac{n}{2} \times \frac{n}{2}$, sending $A_{i,k}$ and $B_{k,j}$ to $O_{i,k,j}$.

\item Recursively partition each octant $(\alpha-1) \lg n$ more times, yielding $n^{3\alpha-6}$ subcubes of size $n^{6-2\alpha}$, each holding 2 submatrices of size $n^{3-\alpha} \times n^{3-\alpha}$.

\item Within each subcube do 3-d stepwise simulation of 2-d systolic matrix multiplication.

\item Recursively add the submatrix results.
\end{enumerate}
\smallskip

\center{Algorithm A: General matrix multiplication in 3-space}

\hrulefill
\end{figure}

Note that this is essentially a manager-worker approach with multiple layers of managers, similar to ones used in some parallel dynamic programming algorithms.
In Algorithm A the $O_{i,j,k}$ in step 1 act as high-level managers and each iteration of step 2 creates exponentially more lower-level managers with smaller problems to work on.
At the lowest level, step 3 assigns each task to a team of workers.
Step 4 is the reverse process of the results at a lower level being combined and then passed upward to a higher level.
In general manager-worker problems the tasks to be worked on may be variable length, or not known in advance, while
here everything is known in advance, greatly simplifying the algorithm.
The problems assigned to each manager, the problems it assigns to its subordinates, and at the bottom layer the tasks performed by the workers, are all divided evenly, perfectly parallelizing the work.
The proof shows that the communication distance, hence communication time, can be optimized as well and is bounded by the time to do the work.
Not only is the distance optimized, but also the bandwidth is large enough to move the entire subarray in time proportional to the time move a single item that far.

Following the recursive approach used by Strassen, we know that if there is a serial algorithm $\mathcal{M}$ and $m$ such that multiplication of $m \times m$ matrices over a ring can be done in $< m^{\alpha}$ termwise multiplications and some fixed number of additions (the number is irrelevant for the recursive analysis) then there is a recursive serial algorithm taking $O(n^\alpha)$ time.
The approach yields a time recurrence of the form $T(n) = a T(n/b) + f(n)$, where $\log_b a = \alpha$ and $f(n) = O(n^{\alpha - \epsilon})$ for some $\epsilon > 0$.
We want an $\alpha < 2\frac{2}{3}$, and the first author to achieve this was Schönhage~\cite{EarliestFastEnough}, who gave an $\alpha < 2.522$, where $f$ also has the desired parallelization.
Reductions in the smallest value of $\alpha$ known are continuing
but so far any improvements in the 21$^{\mathrm{st}}$ century have been tiny and require galactic algorithms.
Further, they are not needed in 3-d.

\begin{figure}[t]
\noindent Matrix multiplication of $n \times n$ matrices $C = A \times B$\\
in a 3-d mesh of size $n^2$, where the operations are over a ring.\\
See the text for an explanation of $m$, $r$ and $s$.
\medskip

For each of the $s$ steps, these are the operations a subcube at the start of a step performs when submatrices arrive from the previous step
\begin{enumerate}
\item Sequentially go through the recursion tree corresponding to $r$ levels of recursion in a breadth-first manner, forming $m^r$ subproblems
\item At each iteration within a step
\begin{enumerate}
\item send one subproblem to each of the children.
\item receive the result from each child
\item combine the results to form answers in the recursion tree
\end{enumerate}
\item When done, return the result to the parent.

\end{enumerate}
\smallskip

\center{Algorithm B: Matrix multiplication over a ring in 3-space}

\hrulefill
\end{figure}

\begin{theorem} \label{thm:3dringmult}
Given two $n \times n$ matrices over a ring in a 3-d mesh-connected computer of size $n^2$, their product can be computed in $\Theta(n^{2/3})$ time, which is optimal over 3-d meshes of any size.
\end{theorem}
\noindent \textit{Proof:}~
Like Algorithm A, multiplications of submatrices will be recursively assigned to smaller subcubes. 
However, it is more complex because the space needs to be kept at $\Theta(n^2)$ to achieve the lower bound given by the communication lower bound noted by Gentleman.

The approach is to have a sequence of $s$ \textit{steps} where at step $r$  levels of recursion are done before subproblems are passed on to the next step.
$r$ will be a constant, and $s$ will be an increasing function of $n$.
The steps are indexed by their ordering in the depths of the recursion.
Each step is a collection of subcubes of the same size and depth in the recursion, with the same number of children, increasing as the depth increases.
Moving from step to step is done in a breadth-first manner, while computations within a step are done in a depth-first manner.
The process is systolic, maintaining a constant flow of problems.
Similar approaches, but not obeying the physics constraints and hence having significantly simpler details,
appear in papers such as~\cite{BallardCommunicationBounds}.

To make the use of steps clear, and to insure that processors are used in only 1 step, we implement the recursion is a slightly unusual manner.
Let $a, b > 1 $ be such that there is a serial recursive matrix multiplication algorithm of $n\times n$ matrices obeys the serial time equation $T(n) = a T(n/b) + f(n)$, where $f(n) = \Theta(n^2)$ (i.e., is like Strassen's algorithm, where a problem is broken into $a=7$ subproblems of size $n/b$, $b=2$).
Thus $T(n) = n^\alpha$, where  $\alpha = \log_b a$.
Choose an algorithm such that $\alpha < 2\frac{2}{3}$.
We assume that $f(n) = O(n^\alpha)$ on a mesh of size $n^2$.

Within a step, each level of recursion increases the volume by a factor of $a/b$, so $r$ levels of recursion has increased it by $(a/b)^r$.
To keep the number of processors the same, the ones working on this step will simulate $(a/b)^r$ processors.
As will be shown below, processors in one step don't work in any others, so none of them need to simulate more than this constant (recall that $r$ is a constant).

When going from one step to the next each subcube sends subproblems to all the children simultaneously, sent as a series of waves.
The number of waves is the same for each level of the recursion.
When subproblems of size $X$ arrive at a subcube in step $i$, $r$ levels are recursion are performed and the resulting subproblems of size $X/m^r$ are passed to the step below it (step $i+1$). 
Further, $r$ will be such that $m^r \geq 8$.
The subcubes at step $i+1$ will each start with multiple subcubes of size $\lfloor(X/m^r)^{1/3}\rfloor$.

Within a step the subproblems are generated the same way, but for each subcube at the start of the step a depth-first search is used,
where the depth reached is the start of the next step.
Being depth-first, the storage at each level is reducing geometrically, so the recursion stack is at most a constant times the original storage.

Leaf blocks of the overall recursion consist of $m$ processors that multiply scalars, using a simple non-recursive algorithm
 such as a 3-d simulation of 2-d systolic algorithm.

To obey the physical constraints and still achieve $\Theta(n^{2/3})$ time, recursively send subproblems to children that are smaller subcubes, where here a child's subcube is inside the parent's, and processors are not in more than one subcube in the sense that no two subcubes are using a processor to do work in each subcube, but the processor may be helping communication within a subcube and between a cube and a child subcube.
Specifically, initially consider a 3-d grid where each dimension has values $0 \ldots \lceil 2n^{2/3} \rceil-1$.
The set of grid points where the binary values of each coordinate end with ``0'' form a cube of size $n^2$, where each such point has distance 2 from its 6 nearest neighbors (points along the boundary have fewer neighbors, and the cube will be slightly larger than $n^2$ if $n^{2/3}$ is not an integer).
Call this the grid of the $0^\mathrm{th}$ step, i.e., the original mesh.
The set of grid points where the binary values of each coordinate end with ``01'' form  a subcube of size $n^2/8$, where each is at distance $4$ from its nearest neighbors.
Call this the grid of the $1^\mathrm{st}$ step down.
The set of grid points with binary values of each coordinate ending with ``011'' form a subcube of size $n^2/64$, where each is at distance $8$ from its nearest neighbors.
This is the grid of the $2^\mathrm{nd}$ step down, \ldots.

The 0th step needs to be proportional to the size of the original mesh, while at the lowest step there must be enough processors there to be able to perform all the pairwise scalar multiplications required.
If there are $s$ steps then the lowest has a grid of size $n^2/8^s$ and
and nearest neighbors are at distance $2^{s+1}$.
To compute all of the pairwise multiplications needed in $\Theta(n^{2/3})$ time we need all of these processors to work at a constant rate, waiting $2^{s+1}$ time steps between multiplications.
I.e., we need $(n^2 \times n^{2/3})/64 \geq \Theta(n^\alpha) = \Theta(n^{\log_b a})$.
This holds if $n^{8/3}/64^s \geq n^\alpha$, i.e., if $n^{8/3-\alpha} \geq 64^s$.
Letting $\delta = \frac{8}{3}-\alpha$, we must have $\delta \log_{64} n \geq s$.

Meanwhile, we need to have created terminal tasks for these processors, i.e., we need to do $\log_b n$ levels of recursion.
This can be done if we do $ r = \log_b n / \delta \log_{64} n$ levels of recursion per step.
This is a constant, so we do this many levels of recursion at one step and then pass the subproblems down to the next step.
Each cube at one step generates 64 subcubes at lower step, each of which has an extent (in its numbering) in each dimension 1/2  that of the upper subcube. 
While the subproblems are smaller in the number of processors in each direction, the beginning iteration of the lower step takes as long as the lowest blocks in the previous iteration.
However, going through $r$ iteration levels means that the blocks at the end of the lower step are significant faster, and this is passed on.

To determine when the lowest step starts working on problems, note that the 0th step takes $\Theta(n^{2/3})$ time to do the first level of recursion.
Then it does the next level, the next, etc., until it starts passing subproblems to the step below.
Each recursive level increases the total work that needs to be done, but it is only increased by a constant proportion of the initial size.
The recursive calls are done in a breadth first manner, where processors have to save many (but with a fixed bound $r$) intermediate results before passing subproblems to the step below.
In $\Theta(n^{2/3})$ time the next step down starts on subproblems where the edge of the subcubes in has decreased by at least a factor of 2..
Thus the lowest step starts doing the pairwise scalar multiplications in $\Theta(n^{2/3})$ time.
The processors there work at a constant rate, which the above construction shows continues for $O(n^{2/3})$ time.
While the steps are passing subproblems down, they answers are also being returned, again taking $\Theta(n^{2/3})$ time, so the algorithm finishes in $\Theta(n^{2/3})$ time, as claimed.

A proof of optimality was given at the start of this section.
$\Box$
\smallskip

There are other matrix problems that have been shown to have the same time complexity as matrix multiplication~\cite{Strassen}, including matrix inversion and determinant, and possibly these too can be performed in $\Theta(n^{2/3})$ time.

\section{Graphs and Paths} \label{sec:graph}

Given the adjacency or weight matrix of a graph of $n$ vertices, using a 2-d mesh to determine various properties of the graph has long been studied.
There are a class of \textit{path-like} problems that are quite similar.
This includes transitive closure, all-pairs shortest path, and all-pairs bottleneck path (also known as the widest path problem) which finds a path which maximizes the smallest weight on the path.
They can all be solved by using F-W matrix multiplication.

The path-like problems can also be solved by repeated squaring using general $(\oplus,\otimes)$ matrix multiplication.
This is the approach we use on a 3-d mesh.
Using general matrix multiplication, Algorithm A shows:

\begin{corollary} \label{cor:3dsquaring}
For any $2 \leq \alpha \leq 9/4$, given the adjacency or weight matrix of a directed graph of $n$ vertices in a 3-d mesh-connected computer of size $n^\alpha$, in $\Theta(n^{3-\alpha} \log n)$ time one can solve the all-pairs shortest path (APSP) and all-pairs bottleneck path problems.
In particular, when $\alpha = 9/4$ the time is $\Theta(n^{3/4} \log n)$. $\Box$
\end{corollary}

This isn't a new approach, in a class on serial algorithms repeated squaring is often mentioned, then they are shown that faster approaches.
Transitive closure can be solved the same way, but has a faster solution by noticing that squaring the adjacency matrix as if it was over a ring results in all nonzero (hence positive) entries corresponding to paths of length 2.
Here one first converts ``true'' to 1 and ``false'' to 0.
Extending to paths of length of powers of two can be achieved using repeated squaring.
To include paths of all lengths first make the diagonal entries 1, then do repeated squaring.
This is a well known approach which works well in our model.

\begin{corollary} \label{cor:transclosure}
Given the adjacency matrix of a directed graph of $n$ vertices in a 3-d mesh of size $n^2$, the transitive closure matrix can be determined in $\Theta(n^{2/3} \log n)$. $\Box$
\end{corollary}

There are several graph problems that can be solved in the same time using the same toolbox.
Examples include finding all bridge edges and all articulation points.

\begin{table*}
\begin{center}

\begin{tabular}{|l|l|l|} \hline
Problem & 2-d:  ~all times $\Theta(n)$   & 3-d replacement:~ ~time, mesh size, algorithm used \\ 
               & mesh size $\Theta(n^2)$ &  \\  \hline
Min spanning tree, & Recursive divide-& $\Theta(n^{2/3})$,~ $\Theta(n^2)$~~~~Same approach \cite{QFSMST84} \\
Component labeling          &  and-conquer \cite{QFSMST84}     &   \\ \hline
 Transitive  &  Warshall's algorithm & $\Theta(n^{2/3} \log n)$,~$\Theta(n^2)$~~~~Square adjacency matrix $\lceil \log_2 n\rceil$ \\
 closure     &   \cite{VanScoy}$^*$             & \hspace*{0.6in} times using (+,*) mult.  (Corollary~\ref{cor:transclosure}) \\ \hline
 All Pairs    &  Floyd-Warshall                    & $\Theta(n^{3/4} \log n)$,~ $\Theta(n^{9/4})$~~~~~Square distance matrix $\lceil \log_2 n\rceil$\\
 Shortest Path (APSP) &  \cite{VanScoy}      &  \hspace*{0.6in} times using (min,+) general mult.\   (Corollary~\ref{cor:3dsquaring}) \\ \hline                                  
 2-d maze        &  Shrinking~\cite{Beyer,Levialdi}$^*$, a & $\Theta(n^{2/3})$,~ $\Theta(n^2)$~~~~~Component labeling \cite{QFSMST84}\\
 reachability & topological operation     & \hspace*{0.6in} \\ \hline                  
 2-d maze   & Recursive APSP & $\Theta(n^{3/4} \log n)$,~ $\Theta(n^{9/4})$~~~~~Same approach\\
 shortest path          &  of subsquares  \cite{MiStPAMI}&   \\ \hline

\end{tabular}
 \smallskip
 
All inputs are $n \times n$ matrices except MST and component labeling which are $n^2$ edges in arbitrary order\\
$^*$ References show how to achieve this using weaker cellular automaton model

\end{center}

\vspace*{-0.15in}

\caption{Replacing $\Theta(n)$ algorithms on 2-d $n \times n$ mesh (hence size $\Theta(n^2)$) with faster ones in 3-d}    \label{tab:Replace2d}

\hrulefill
\end{table*}

\section{Mazes} 

Mazes are a grid of black/white pixels (in 2-d) or voxels (in 3-d), where 2 whites are considered adjacent if they share an edge or face, respectively
(for the Jordan curve theorem to hold in 2-d, black adjacency includes corners, and in 3-d adjacency includes edges).
To simplify description, the terminology for 2-d mazes will be used for both.
Throughout mazes are $n \times n$, or $n \times n \times n$, respectively, stored one pixel per processors.

One white pixel is labeled ``start'' and another white pixel is labeled ``finish'',
and a \text{path} from start to finish is a sequence of adjacent white pixels.
Maze problems have been analyzed for decades, with two primary problems of concern: deciding if there is such a path (aka reachability or connectedness), and, if there is such a path, marking the pixels on a shortest one.
Minsky and Papert famously showed that determining if there is such a path in a 2-d maze cannot be decided by a 1-level perceptron~\cite{perceptrons}.

\subsection{2-Dimensional Maze} \label{sec:2dmaze}

Beyer~\cite{Beyer} and Levialdi~\cite{Levialdi} showed that a 2-d $n \times n$ cellular automaton could decide if the start and finish are connected in $\Theta(n)$ time.
They used
a simple, elegant, shrinking procedure which preserves connectivity, where there is a path from start to finish if and only if the start and finish are ultimately brought together.
However, shrinking does not preserve distance, and Beyer raised the question of the minimum time required by a 2-d cellular automaton to mark a shortest path, or even to just mark a simple path.
These questions from the 60s are still open.

There is a simple, but quite slow, algorithm: have the start send a signal to its white neighbors, they pass it on to their neighbors, and so on.
Each white pixel records which edge it first received the signal from (ties are broken arbitrarily).
When the finish receives the signal (if it ever does) these records are used in reverse order to mark the path.
This breadth-first ``wave propagation'' or ``diffusion'' takes time linear in the  path length, which can be $\Theta(n^2)$.

Surprisingly there is continuing interest in using this approach:
special purpose circuits have been built~\cite{wavecircuit}, memristor~\cite{memristor} and fluidic~\cite{fluidic} networks have been used, fatty acid chemistry~\cite{fattyacid}
and even slime molds~\cite{slimemold} have been used.

By switching from a 2-d cellular automaton to a 2-d mesh-connected computer one can do better.
Partition the maze into quadrants, and in each quadrant the white boundary pixels, and the start or finish pixels if they are in the quadrant, are viewed as vertices in a graph.
Suppose the all-pairs shortest distance has been computed for these vertices.
In particular, the distance from the start and end vertices to every boundary pixel of their quadrant has been determined.
Take the union of these vertices for all quadrants, compute the all-pairs shortest distance for them, and for each pair record the first step on an optimal path.
Note that the square has enough processors to hold the distance matrix.
If this square is a subsquare of a larger one then the vertices along its interior edges are removed and the process iterates.
Using the shortest path algorithm mentioned in Section~\ref{sec:graph}, the total time is $\Theta(n)$.
See~\cite{MiStPAMI} for further details.

\subsection{3-Dimensional Maze} \label{sec:3dmaze}

Beyer~\cite{Beyer} and Levialdi's~\cite{Levialdi} 2-d reachability algorithms are based on homotopy, shrinking white regions without ever combining two disjoint ones.
In 3-space, however, such a homotopy procedure cannot be used. For example, consider adjacent links in a chain. It is impossible to shrink one link without ultimately coming in contact with the other.
Beyer raised the question of the time required for reachability in a 3-d maze, a question repeated in Minsky and Papert~\cite{perceptrons}.
By using a graph-based connected components approach a 3-d cellular automaton can determine reachability in $\Theta(n)$ time~\cite{QFSmaze} for an $n \times n \times n$ maze.

All of the results mentioned are achieved by simple cellular automata, but even when the processors are the more powerful mesh processors there are no known algorithms to produce a simple path, let alone a shortest path, from start to goal in $o(n^3)$ time for 3-d.
A shortest simple path can be found in $\Theta(n^2)$, and $\Theta(n^3)$ time, respectively for 2-d and 3-d, by cellular automata using simple breadth-first search.
The optimal times to find either any simple path or any shortest path, for 2-d or 3-d using cellular automata or 3-d for mesh processors, are all open questions when the mesh and maze dimensions must be the same.

For a 3-d mesh, the recursive approach used above for 2-d mazes computes a distance matrix for a slice through the middle.
In an $n \times n \times n$ maze there are $\Theta(n^2)$ elements in a slice through the middle, so the matrix requires $\Theta(n^4)$ space, which exceeds the input size.
Expanding to a 3-d mesh of size $\Theta(n^4)$ and simulating the 2-d shortest path algorithm takes time $\Theta(n^2)$, and it is easy to show that this is the total time (the time for paths on the subcubes decreases geometrically).

One can do better by replacing the 2-d shortest path algorithm with repeated squaring on a yet larger mesh.
Using Corollary~\ref{cor:3dsquaring}, one has:
\vspace*{-0.05in}

\begin{theorem} \label{thm:3dmaze}
Given a solvable $n \times n \times n$ maze in a 3-d mesh of size $n^c$, $4 \leq c \leq 9/2$, a shortest path from start to finish can be marked in $\Theta(n^{6-c} \log n)$ time.
In particular, when $c=9/2$ the problem can be solved in $\Theta(n^{3/2} \log n)$ time.
\end{theorem}

\section{Final Remarks}  \label{sec:final}

3-dimensional meshes basically represent the limits classical physics imposes on computation in 3-space.
The constants depend on the technology, but words have a minimum volume and, for fixed power, take unit time to travel unit distance.
Technically the results here are still only an approximation in that we treated processors' memory as constant volume when we actually used a fixed number of words of $\Omega(\log n)$ bits, and we assumed that scalar operations took constant time.
In some cases it may be that if the input matrices are $n\times n$ with $g(n)$ size entries, i.e., the real input size is $\Theta(n^2 g(n))$, then the problem can be solved in $\Theta(n^\alpha g(n))$ serially and on the mesh in $\Theta((n f(n))^{2/3})$ time, but this makes an assumption about the parallelization of $f$.
Even when it is true it would probably require using clerks~\cite{QFSmaze} or some equivalent.

There are many papers on parallel matrix multiplication, a small sample of which is \cite{Agarwaletal3D,SUMMA,FaultTolSUMMAMatrixMult,RedBlueMatrixMult,Schatzetal3Dmedium,2.5DCommunicationOptimal,Suel,ThKuSort}.
Some of them utilize aspects of dimension, such as the 2.5D and 3D variants, though often these have a somewhat different interpretation of dimension. 
Some use Strassen's algorithm, which is practical as opposed to the serial matrix algorithm that formed the basis of our parallel algorithm for rings, and some minimize the communication that any single processor requires.
Many of these are optimized by having a nontrivial amount of memory per processor.
However, none of them take into account that information takes time linear in distance to be moved, instead analyzing in terms of access in RAM taking constant time, and that the time for passing information between boards depends only on the amount being transferred. 
Nor do they take into account that the presence of wires increases the distance information must travel.
Some, for example, have ``a link in the network between each pair of processors"~\cite{BallardDemmelStrassen}, and some have communication difficulties (``fast matrix multiplication algorithms (e.g., Strassen’s) running on a 3D torus will suffer from contention bottlenecks''~\cite{BallardCommunicationBounds}).
These may be reasonable assumptions for algorithmic analyses of computers of practical size, but we've concentrated on fundamental limits as the size goes to infinity,
The primary contribution of this paper is to show that on a fine-grained model, with only a constant amount if memory per processor, all of this can be taken into account and still accomplish matrix multiplication in $\Theta(n^{2/3})$ time, achieving the minimum possible when physical constraints are taken into account, though the implied constants are far larger than the implementations available.

Interest in 3-d layouts goes back at least to the 80's~\cite{LeRo3dcircuit,Rosenberg3DIC}, and was revived in the early 2000's~\cite{DARPA3D,Pavidis3D,Topol3D}.
Because other techniques for speeding up computation and reducing energy consumption are petering out, there is rapidly increasing interest in 3-dimensional chips and modules, especially for exascale computing~~\cite{DARPA3D,DOE:SciGrandChal,DARPAExascale} and now for zetascale.
Faster matrix multiplication and similar operations might also speed up machine learning.
Currently the commercial chips for neural nets use systolic 2-d algorithms~\cite{Amazon,Google} but there is interest in far more complex problems which require much more powerful hardware.

Despite this interest in hardware, there has been far less in fundamental 3-d algorithms.
We examined some basic problems on fine-grained systems where the solutions in 3 dimensions are quite different from those in 2 dimensions.
In 2-d, the fastest possible (and achieved) sorting, general matrix multiplication, all-pairs shortest paths in a graph, and shortest paths in a maze all take time linear in the edge length.
In 3-d the edge length is smaller than linear time speedup of $\Theta(n^3)$ algorithms, and this difference greatly enlarges the algorithmic possibilities.
Suddenly expanding to a larger mesh and sub-cubic matrix multiplication become useful tools.

For 3-d fine-grained models there are numerous open questions, e.g., can the logarithmic factors be eliminated in the 3-d algorithms for path-like problems (Sec.~\ref{sec:graph})?
Solving 2-d mazes have a sharp change in the time when going from a 3-d mesh the same size as the maze ($\Theta(n^{2/3} \log n)$ time) to one just large enough to hold the all-pairs shortest distance matrix of a slice through the center ($\Theta(n^{4/3} \log n)$ time).
What about intermediate sizes?     

There are also questions concerning other path problems.
Here we could have required that $\alpha$ be the smallest constant such that matrix multiplication over a ring can be performed in $O(n^\alpha)$ time (i.e., the constant commonly called $\omega$), but that was not necessary.
Williams~\cite{RWilliamsAPSP} gave a $\Theta(n^3/2^{\omega (\log n)^{1/2}})$ serial algorithm for the all-pairs shortest path problem, but it is unlikely that this improvement can be exploited on a mesh.
All-pairs bottleneck paths has more potential since Duan and Pettie~\cite{DuanMaxMinMatrix} gave a serial algorithm taking $\Theta(n^{(\omega+3)/2})$ time, and the variant with vertex weights, not edge weights, has an even faster algorithm~\cite{APBPVertex}.
Williams and Williams have shown tight connections between various matrix related problems~\cite{WilliamsSubcubic}, ones where the approaches herein may be useful.

Finally, there is the question of whether 3-d meshes are feasible.
While companies and researchers have stacked chips for some communication in the 3rd dimension, the 3rd dimension is always a very weak link.
More interesting, but far more futuristic, is the possibility of using a biological framework to construct fine-grained computational entities \cite{HighPerfBiocomputing,CellularSupremacy,ComputingToLife,MolecularMotor,FutureComputers,wikiBiologicalCompute}.
What they could reasonably compute is a significant research question, but there are numerous examples of biological creatures that live in 3-d and are able to compute.

\end{document}